# Ultra-high-Q racetrack micro-ring based on silicon nitride

SHUAI CUI,[1,2] KAIXIANG CAO,[1,2] YUAN YU [1,2*] AND XINLIANG ZHANG[1,2]

[1]*Wuhan National Laboratory for Optoelectronics and School of Optical and Electronic Information, Huazhong University of Science and Technology, Wuhan 430074, China*
[2]*Optics Valley Laboratory, Wuhan, 430074, China*
**yuan_yu@hust.edu.cn*

**Abstract:** Ultra-high-Q resonators are fundamentally important to optics and microwave photonics. Up to now, it is still very challenging to boost the Q factor while maintaining a compact size for a resonator. Herein, we proposed and demonstrated an ultra-high-Q silicon-nitride ($Si_3N_4$) racetrack resonator with uniform multi-mode $Si_3N_4$ photonic waveguides. It consists of two multi-mode straight waveguides connected by two multi-mode waveguide bends (MWBs). In particular, the MWBs are based on modified Euler curves, and a multi-mode straight waveguide directional coupler is used for the fundamental mode coupling and avoid exciting higher-order modes in the racetrack. In this way, the fundamental mode is excited and propagates in the multi-mode racetrack resonator with ultra-low loss. Meanwhile, it helps to achieve a compact 180° bend to reduce the chip footprint. In this paper, the propagation loss of the fundamental mode is significantly reduced with standard fabrication process by broadening the waveguides width to as wide as 3 μm. Results show that an ultra-high-Q resonator with an intrinsic Q of TE mode is $4.57×10^7$, and the corresponding propagation loss of the waveguide is only 1.80 dB/m. To the best of our knowledge, this is the highest Q value of the ring resonator with only 2.226 mm ring length reported so far. The proposed ultra-high-Q $Si_3N_4$ resonator can be used to microwave photonic filters and optoelectronic oscillators with large operation bandwidth.

## 1. Introduction

The past decades have witnessed dramatic progress in the development of photonic integrated circuits in silicon-on-insulator (SOI). On one hand, the key advantages of silicon (Si) photonics include large-area substrates, mature and high-yield CMOS compatible fabrication, and availability of sophisticated assembly processes. These features, together with the high refractive index contrast in Si photonic platforms, can enable the manufacturing of densely integrated electronic and photonic components at low costs and high volumes [1-3]. On the other hand, although the Si waveguide has high nonlinearities, but suffer from two-photon absorption at telecommunication wavelengths, drastically restricting its applications in nonlinear photonics. Recently, the $Si_3N_4$ [4,5] has emerged as an attractive CMOS compatible alternative, as it offers low loss and high nonlinearity. Meanwhile, the $Si_3N_4$ can avoid two-photon absorption over the telecommunication wavelengths, which can further reduce the waveguide loss. The lower index contrast of $Si_3N_4$ (n ≈ 2) waveguides with silica ($SiO_2$, n ≈ 1.45) cladding compared to Si (n ≈ 3.48) waveguides also reduces the waveguide loss caused by sidewall roughness scattering. Because of these advantages of $Si_3N_4$ waveguides, it has been proposed to integrate low-loss $Si_3N_4$ waveguides with other active and passive optical components [6-11] to enhancing the device performance. Exploiting the low-loss $Si_3N_4$ waveguide to boost the device performance has been attracting significant interest.

As is known, the propagation loss mainly originates from the scattering at rough sidewalls. Much effort has been made to reduce the loss of optical waveguides by improving the fabrication processes. To reduce the propagation loss, some special fabrication processes have been developed to smoothen the waveguide sidewall [12-14]. However, these special fabrication techniques are not standard process and incompatible with those standard processes in foundries [15]. Moreover, these fabrication approaches are not available generally for various material platforms. Therefore, it is still greatly desired to achieve low-loss optical waveguides by using regular standard fabrication processes. A potential way is to weaken the optical field intensity at the boundaries of a photonic waveguide, so that light scattering induced by the sidewall roughness is reduced significantly. Further, the Q factor of fabricated micro-ring resonators can be increased. For example, the propagation loss of silicon waveguide can be reduced to 0.21dB/cm by combining multi-mode waveguide and single-mode waveguide, which is used to ensure single-mode transmission in the micro-ring resonator [16]. To further reduce the propagation loss, the single-mode waveguide is also replaced by multi-mode waveguide, which is designed with Euler bend to ensure single-mode transmission. By doing so, the intrinsic Q factor is increased to $1.02 \times 10^7$ [17]. Notably, $Si_3N_4$ waveguides can achieve much lower propagation loss than silicon waveguides [18,19]. Therefore, micro resonators based on $Si_3N_4$ waveguide can achieve much higher Q values [20]. However, the lower index contrast between $Si_3N_4$ core and $SiO_2$ cladding push the perimeter of high Q resonators to be as large as 9.65 mm [21], which significantly reduces the compactness and induces a small free sepctral range (FSR). So far, to achieve a compact ultra-high-Q resonator based on $Si_3N_4$ waveguide is still very challenging.

In this paper, an ultra-low-propagation-loss $Si_3N_4$ photonic waveguide of 1.80 dB/m is achieved by optimizing optical waveguide. The waveguide is fabricated based on standard fabrication process. With the help of Euler-bends, adiabatic propagation of fundamental mode in multi-mode waveguides is ensured, and high-order modes are suppressed successfully. The loaded and intrinsic Q factors of the fabricated $Si_3N_4$ micro-ring resonator are as high as $1.55 \times 10^7$ and $4.57 \times 10^7$, respectively. Thanks to the Euler bend designed in the resonator, the perimeter of ring resonator is only 2.226 mm, and the experimentally measured FSRs of the TE and the TM modes are around 0.52 and 0.51 nm, respectively. The proposed ultra-high-Q micro-ring resonator with large FSR is very promising in microwave photonics, classical and quantum information processing and ions-sensing applications.

## 2. Device design

Fig. 1(a) and (b) show the three-dimensional (3D) and the top views of the designed ultra-high-Q racetrack resonator based on uniform multi-mode $Si_3N_4$ waveguide. The racetrack micro-ring resonator is composed of two multi-mode straight waveguides and two multi-mode waveguide bends (MWBs) based on modified Euler curves. A directional coupler is used to couple the light from the bus waveguide to the micro-ring. Notably, the directional coupler is realized by using two multi-mode straight waveguides with the same width. By controlling the coupling length and the coupling gap, the coupling only occurs in the multi-mode waveguide, which achieved sufficient coupling coefficient for fundamental mode and suppress high-order mode excitation.

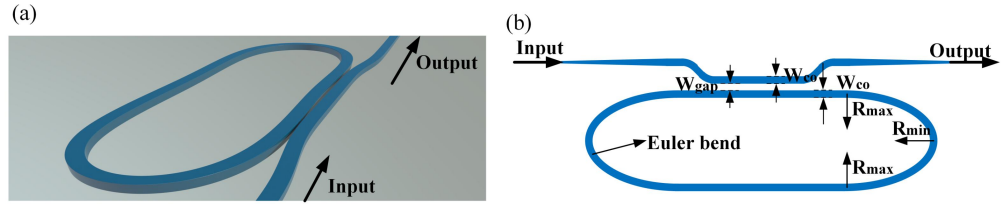

**Fig. 1.** Schematic configurations of the proposed ultra-high-Q micro-ring resonator.

(a) 3D view; (b) Top view.

Fig. 2(a) shows the cross section of the $Si_3N_4$ waveguide, whose core width $W_{co}$ is beyond the single-mode regime in order to weaken the light-sidewall interaction and thus reduce the light scattering loss at the sidewalls. The interaction between the optical field and the roughness is responsible for distributed scattering loss. Here we use a $n-w$ model to analyze the scattering loss, which provide a comprehensive analysis for the fundamental role played by the sensitivity of the effective index ($n_{eff}$) of the optical mode to waveguide width ($w$) variations [22]. This approach enables an accurate description of practical optical waveguides and provides simple design rules for optimizing the waveguide geometry, thus to reduce the scattering loss generated by sidewall roughness. According to the Si$_3$N$_4$ design rules (LIGENTEC, Switzerland), for a single mode Si$_3$N$_4$ waveguide with 1 μm wide and 800 nm thick core region, the scattering loss caused by the sidewalls is approximate 10 dB/m [23]. The relationship between waveguide width and transmission loss can be predicted by using the $n-w$ model. From Fig. 2(b), it can be seen that when the average deviation σ is determined, the transmission loss decreases with the increase of waveguide core width $W_{co}$. This happens because when the waveguide core becomes wider, the optical field on the side wall of the waveguide becomes weaker. We can also see that when $W_{co}$ is larger than 3 μm, the transmission loss is not very sensitive to the increase of $W_{co}$. Further, as the waveguide width increases, the MRR footprint will increase significantly. Therefore, to balance the bending radius and the waveguide loss, the waveguide width is designed to be 3 μm. We simulated the mode field in a waveguide of 3 μm width and found that the lowest four modes (TE0, TE1, TE2 and TE3) are all well supported. So in wider waveguides, we need to design a longer coupling length and a smaller gap of the MRR carefully, aiming to get a sufficient coupling ratio for the fundamental mode and to avoid exciting high order modes as much as possible. Accordingly, when it is assumed that σ = 2.5 nm and the correlation length of Lc is 50 nm, the estimated transmission loss of the Si$_3$N$_4$ waveguide is 1.2 dB/m. Therefore, the Q value of the micro-ring resonator can be significantly improved.

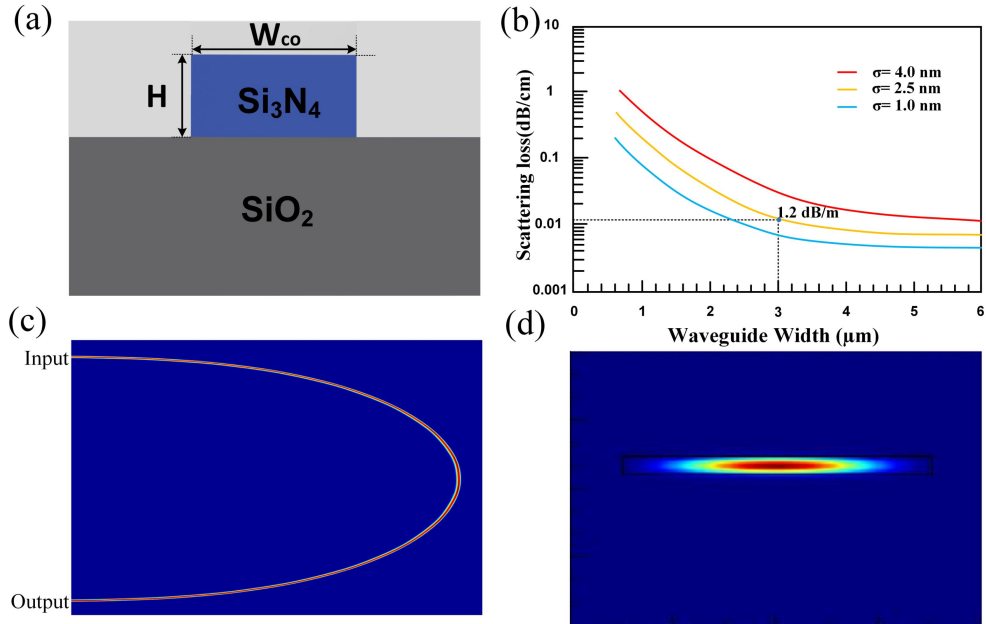

**Fig. 2.** Design of the ultra-high-Q micro-racetrack resonator. (a) Cross section of the Si$_3$N$_4$ optical waveguide of the resonator. (b) The relationship between the transmission loss and the waveguide core width based on $n-w$ model. (c) Simulated light propagation in the designed 180° Euler MWB. (d) Simulated mode profile at the output port of the bus waveguide.

The directional coupler based on multi-mode waveguides should be carefully designed to ensure adiabatic coupling for the fundamental mode. The gap width $w_{gap}$ and the length $l_0$ are optimized to avoid excitation of high-order modes. Meanwhile, the micro-ring resonator is designed to be under coupled to further reduce the coupling loss. In addition to the adiabatic coupling, the adiabatic transmission of the fundamental mode in the ring should also be ensured. Therefore, the MWB should also be carefully designed in order to avoid the excitation of high-order modes. When the waveguide width is as wide as 3.0 µm, the radius of MWB is usually required to be as large as several millimeters. To achieve high integration density as well as a large free-spectral range (FSR), the MWB with modified Euler curve is used in the micro-ring resonator. The modified Euler curve means that the curvature of the arc has a linear relationship with the arc length. Assuming the curvature radius is varied from the maximal $R_{max}$ to the minimal $R_{min}$, the Euler bend is defined as

$$\frac{d\theta}{dL} = \frac{1}{R} = AL + \frac{1}{R_{max}}, \tag{1}$$

and

$$A = (\frac{1}{R_{min}} - \frac{1}{R_{max}})/L_0, \tag{2}$$

where $\theta$ is the center angle corresponding to the unit arc length, $L$ is the arc length of the curve, $R$ is the radius of curvature of the arc, $A$ is a constant, $L_0$ is the arc length of the quarter Euler curve, and $R_{max}$ and $R_{min}$ are the maximal and minimal radiuses of the Euler curve respectively.

As shown in Fig. 1(b), the 180° bend is realized by a pair of 90° Euler bends. Notably, the maximum $R_{max}$ should be large enough to ensure negligible mode mismatch at the junction between the multi-mode straight waveguide and the MWB, while $R_{min}$ should be chosen to ensure adiabatic transmission. When the waveguide width is 3 µm, the simulated light propagation in the designed of 180° MWB is shown in Fig. 2(c), in which the $R_{max}$ and $R_{min}$ are designed to be 4000 and 100 µm, respectively. In the present design, the power coupling ratio for the designed directional coupler is about 0.005 by choosing $w_{gap}$ = 0.8 µm and the length $l_0$ = 400 µm. From Fig. 2(d), it can be observed that the optical field is well confined in the Si$_3$N$_4$-core region at the output port of the bus waveguide. The perimeter of the whole micro-ring is 2226 µm.

## 3. FABRICATION AND MEASUREMENT

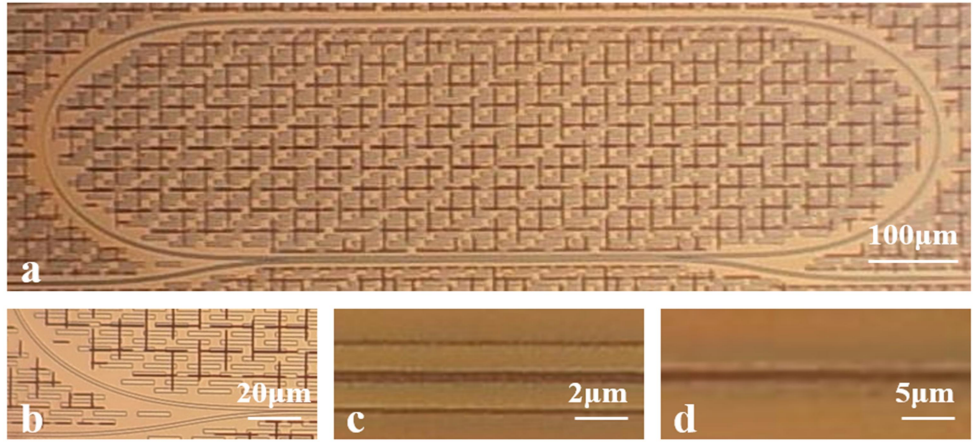

**Fig. 3.** Microscope images of the fabricated ultra-high-Q Si$_3$N$_4$ racetrack resonator. (a) The global view of the fabricated device; (b) The modified Euler bend; (c) Directional coupler; (d) End coupler.

Fig. 3 shows the microscope image of the fabricated ultra-high-Q racetrack resonator (LIGENTEC, Switzerland). Fig. 3(a) shows the global view of the resonator with a compact size of 0.27×0.98 mm$^2$. The microscopic images of the modified Euler-bend, the directional coupler, and the end coupler are shown in Fig. 3(b), (c) and (d), respectively.

Fig. (4) shows the experimental setup for characterizing the Q factor of the fabricated ultra-high-Q $Si_3N_4$ racetrack resonator, which is also used to realize an MPF with ultra-narrow bandwidth. The optical carrier emitted by a tunable laser source (TLS) is input into a phase modulator (PM) via a polarization controller (PC1). Then the phase modulated light is launched into an optical bandpass filter (OBPF). One of the first order sidebands is suppressed by the OBPF and a single sideband (SSB) signal is achieved correspondingly. PC2 is used to adjust the polarization state of the light to the chip. Then the light is amplified by an Erbium Doped Fiber Amplifier (EDFA) and coupled into the chip. Then the SSB is launched into the chip. After filtered by MRR, the output signal is routed to the high speed photodetector (PD) (SHF AG Berlin) with a bandwidth of 40 GHz. Then a microwave photonic notch filter (MPNF) is obtained. Notably, the transmission of the ultra-high-Q resonator is mapped to the transmission of the MPNF. Therefore, the transmission of the ultra-high-Q resonator is measured by vector network analyzer (VNA, Anritsu MS4647B) precisely.

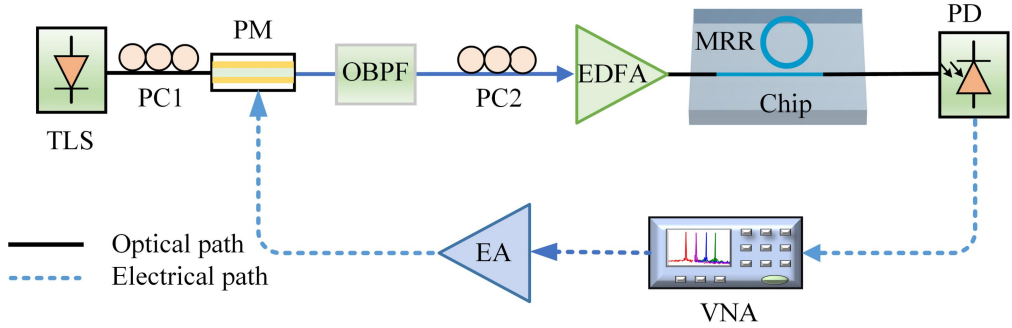

**Fig. 4.** Experimental setup of characterizing the Q factor of the ultra-high-Q $Si_3N_4$ racetrack resonator based on VNA. TLS: tunable laser source; PC: polarization controller; PM: phase modulator; OBPF: optical bandpass filter; EDFA: erbium-doped fiber amplifier; PD: photodetector; VNA: vector network analyzer; EA: electronic amplifier;

Fig. 5(a) shows the measured transmission spectrum at the through port of the resonator. Notably, Fig. 5(a) is obtained by adjusting the wavelength of TLS to enlarge the measurement range. It can be seen that both TE and TM modes exist in an FSR. This is because the thickness of $Si_3N_4$ waveguide is as large as 800 nm, which also supports TM mode transmission. Notably, there are no obvious resonant notches of high order modes, which also proves that the designed micro-ring resonator can ensure adiabatic transmission for the fundamental mode. By using the simulation software MODE Solutions, the group refractive index of $TE_0$ mode is about 2.028 and that of $TM_0$ mode is about 2.064 when the waveguide width is 3 μm and the thickness is 800 nm. The calculated FSRs of $TE_0$ and $TM_0$ modes are 0.53 and 0.52 nm, respectively. The measured FSRs of $TE_0$ and $TM_0$ modes in Fig. 5(a) are 0.52 and 0.51 nm respectively. The difference between the simulated and the measured results lies in the drifts of the laser wavelength and the resonant wavelength during experiment. However, the relative magnitude of the FSRs of $TE_0$ and $TM_0$ modes is not changed. Therefore, the mode with a larger FSR is $TE_0$ mode, while the other is $TM_0$ mode. To measure the resonant notch bandwidth of the $TE_0$ mode, the major resonance normalized transmission of the TE mode is shown in Fig. 5(b). It is shown that the full width at half maximum (FWHM) bandwidth is 0.10 pm, which corresponds to 12.50 MHz. It indicates that the loaded Q of the racetrack resonator is 1.55×10$^7$, and the corresponding intrinsic Q is 4.57

×10$^7$. The calculated propagation loss for the ultra-high-Q racetrack resonator is as low as 1.80 dB/m.

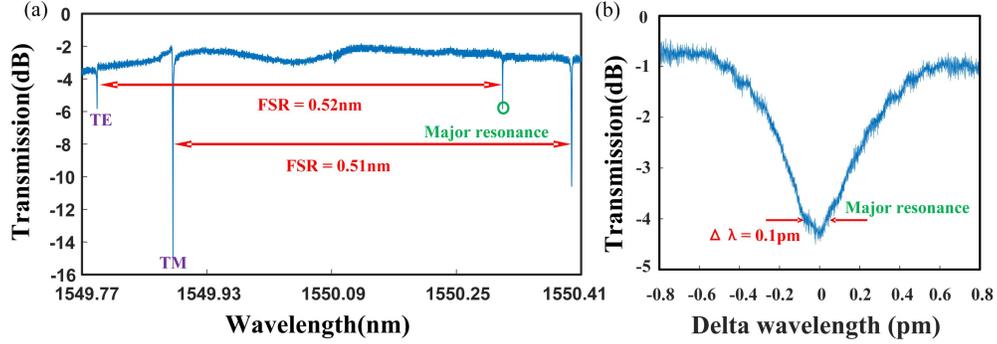

**Fig. 5.** Measured results. (a) Measured transmission of the resonator. (b) Zoomed-in view of major resonance.

When only TE mode is required, we can adjust PC2 to make the state of polarization (SOP) to be aligned with the TE mode of the optical waveguide. Fig. 6(a) shows the measured transmission of the resonator. It can also be observed that there is only TE mode resonant in the micro-ring. Fig. 6(b) shows the resonant notch of TE$_0$. The measured FWHM bandwidth is about 0.28 pm, which corresponds to 35 MHz. Therefore, the loaded Q of the racetrack resonator is 5.5×10$^6$. The extinction ratio (ER) and the loaded Q value of the ring can be changed by adjusting the PC, which can be used in different scenarios. For example, in the case of high Q, the micro-ring resonator can be used for the generation of optical frequency comb. When the ER is adjusted to a larger value, the micro-ring resonator can be applied to microwave photonic filtering.

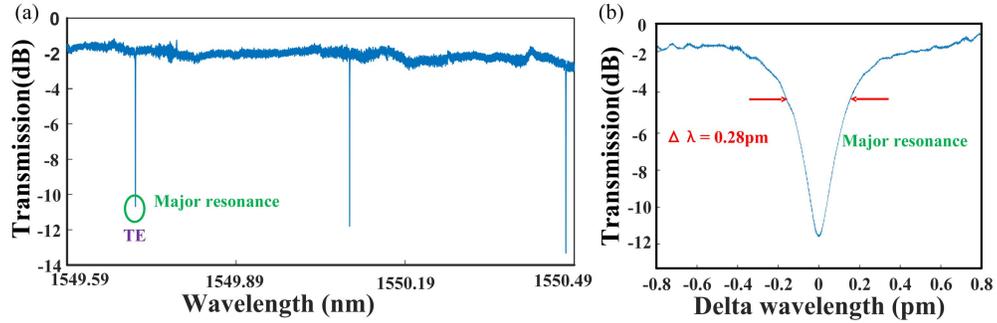

**Fig. 6.** Measured results when only TE mode resonance exists. (a) Measured transmission of the resonator; (b) Zoomed-in view of the major resonance.

**Table 1. Comparison of reported high-Q resonators based on Si$_3$N$_4$**

| Reference | Geometry ($W_{co}$ × H) (μm) | Propagation Loss (dB/m) | Perimeter (mm) | Q (load) | Q (intrinsic) |
|---|---|---|---|---|---|
| [24] | 2.8×0.08 | 2.9 | 12.6 | 1.1×10$^6$ | 7.0×10$^6$ |
| [25] | 2.8×0.10 | 3.0 | 13 | 3.5×10$^6$ | 9.5×10$^6$ |
| [26] | — | 20 | 4.5 | 7.5×10$^5$ | 2.3×10$^6$ |
| [21] | 11.0×0.04 | 0.08 | 61 | 2.6×10$^7$ | 8.1×10$^7$ |
| [27] | 8.0×0.10 | 0.045 | 50 | 1.50×10$^8$ | 2.2×10$^8$ |
| This work | 3.0×0.80 | 1.80 | 2.226 | 1.55×10$^7$ | 4.57×10$^7$ |

Table 1 shows the comparison of recently reported high-Q $Si_3N_4$ resonators with our work. We can see that an ultra-high-Q resonator with compact size is achieved by introducing Euler MWB. The compact size increases the integration density, as well as the FSR of the ring up to 0.52 nm, which is very promising in microwave photonics. Furthermore, our proposed ultra-high-Q resonator is fabricated with standard fabrication processes.

## 4. CONCLUSION

In summary, we have proposed and demonstrated an ultra-high Q racetrack resonator based on $Si_3N_4$ waveguide. Using adiabatic coupling and adiabatic propagation of fundamental mode in multi-mode waveguides, only fundamental mode resonance is realized. In particular, the scattering loss is analyzed by using the $n – w$ mode, and the waveguide width is optimized correspondingly. Meanwhile, a directional coupler based on multi-mode waveguides has been used to achieve adiabatic coupling for the fundamental mode. At the same time, the modified Euler multi-mode bending is employed to reduce the size and avoid exciting high-order modes in the micro-ring simultaneously. The measured FSR of the $TE_0$ mode is 0.52 nm, and the loaded and the intrinsic Q factors of the fabricated $Si_3N_4$ racetrack can reach $1.55\times10^7$ and $4.57\times10^7$, respectively. Therefore, the waveguide loss of the ring resonator is 1.80 dB/m. The proposed approach opens wide opportunities for the applications of $Si_3N_4$ micro-resonators in microwave photonics, classical and quantum information processing and ions-sensing applications.

**Funding**

National Natural Science Foundation of China (61975249, 62005087); National Key Research and Development Program of China (2018YFB2201700, 2018YFA0704403); Open Projects Foundation of Yangtze Optical Fiber and Cable Joint Stock Limited Company (YOFC) (SKLD2006); Program for HUST Academic Frontier Youth Team (2018QYTD08).

**Disclosures**. The authors declare no conflicts of interest.